# Integrating Machine Learning and Multiscale Modeling:
## Perspectives, Challenges, and Opportunities in the Biological, Biomedical, and Behavioral Sciences

Mark Alber, Mathematics, University of California, Riverside
Adrian Buganza Tepole, Mechanical Engineering, Purdue University
William Cannon, Computational Biology, Pacific Northwest National Lab
Suvranu De, Mechanical, Aerospace and Nuclear Engineering, Rensselaer Polytechnic Institute
Salvador Dura-Bernal, Physiology & Pharmacology, State University of New York
Krishna Garikipati, Mechanical Engineering and Mathematics, University of Michigan
George Karniadakis, Division of Applied Mathematics, Brown University
William W. Lytton, Physiology & Pharmacology, State University of New York and Kings County Hospital, Brooklyn
Paris Perdikaris, Mechanical Engineering, University of Pennsylvania
Linda Petzold, Computer Science and Mechanical Engineering, University of California, Santa Barbara
Ellen Kuhl, Mechanical Engineering and Bioengineering, Stanford University

**Abstract.** Fueled by breakthrough technology developments, the biological, biomedical, and behavioral sciences are now collecting more data than ever before. There is a critical need for time- and cost-efficient strategies to analyze and interpret these data to advance human health. The recent rise of machine learning as a powerful technique to integrate multimodality, multifidelity data, and reveal correlations between intertwined phenomena presents a special opportunity in this regard. However, machine learning alone ignores the fundamental laws of physics and can result in ill-posed problems or non-physical solutions. Multiscale modeling is a successful strategy to integrate multiscale, multiphysics data and uncover mechanisms that explain the emergence of function. However, multiscale modeling alone often fails to efficiently combine large data sets from different sources and different levels of resolution. Here we demonstrate that machine learning and multiscale modeling can naturally complement each other to create robust predictive models that integrate the underlying physics to manage ill-posed problems and explore massive design spaces. We review the current literature, highlight applications and opportunities, address open questions, and discuss potential challenges and limitations in four overarching topical areas: ordinary differential equations, partial differential equations, data-driven approaches, and theory-driven approaches. Towards these goals, we leverage expertise in applied mathematics, computer science, computational biology, biophysics, biomechanics, engineering mechanics, experimentation, and medicine. Our multidisciplinary perspective suggests that integrating machine learning and multiscale modeling can provide new insights into disease mechanisms, help identify new targets and treatment strategies, and inform decision making for the benefit of human health.

**MOTIVATION**

Wouldn't it be great to have a virtual replica of ourselves to explore our interaction with the real world in real time? A living, digital representation of ourselves that integrates machine learning and multiscale modeling to continuously learn and dynamically update itself as our environment changes in real life? A virtual mirror of ourselves that allows us to simulate our personal medical history and health condition using data-driven analytical algorithms and theory-driven physical knowledge? These are the objectives of the Digital Twin [Madni et al., 2019]. In health care, a Digital Twin would allow us to improve health, sports, and education by integrating population data with personalized data, all adjusted in real time, on the basis of continuously recorded health and lifestyle parameters from various sources [Buynseels et al., 2018; Liu et al., 2019; Topol, 2019]. But, realistically, how long will it take before we have a Digital Twin by our side? Can we leverage our knowledge of machine learning and multiscale modeling in the biological, biomedical, and behavioral sciences to accelerate developments towards a Digital Twin? Do we already have digital organ models that we could integrate into a full Digital Twin? And what are the challenges, open questions, opportunities, and limitations? Where do we even begin? Fortunately, we do not have to start entirely from scratch. Over the past two decades, multiscale modeling has emerged into a promising tool to build individual organ models by systematically integrating knowledge from the tissue, cellular, and molecular levels, in part fueled by initiatives like the United States Federal Interagency Modeling and Analysis Group IMAG. Depending on the scale of interest, multiscale modeling approaches fall into two categories, ordinary differential equation-based and partial differential equation-based approaches. Within both categories, we can distinguish data-driven and theory-driven machine learning approaches. Here we discuss these four approaches towards developing a Digital Twin.

***Ordinary differential equations characterize the temporal evolution of biological systems***. Ordinary differential equations are widely used to simulate the integral response of a system during development, disease, environmental changes, or pharmaceutical interventions. Systems of ordinary differential equations allow us to explore the dynamic interplay of key characteristic features to understand the sequence of events, the progression of disease, or the timeline of treatment. Applications range from the molecular, cellular, tissue, and organ levels all the way to the population level including immunology to correlate protein-protein interactions and immune response [Rhodes et al., 2018], microbiology to correlate growth rates and bacterial competition, metabolic networks to correlate genome and physiome [Cuperlovic-Culf ,2018; Shaked et al. 2016], neuroscience to correlate protein misfolding to biomarkers of neurodegeneration [Weickenmeier et al., 2018], oncology to correlate perturbations to tumorigenesis [Nazari et al., 2018], and epidemiology to correlate disease spread to public health. In essence, ordinary differential equations are a powerful tool to study the dynamics of biological, biomedical, and behavioral systems in an integral sense, irrespective of the regional distribution of the underlying features.

***Partial differential equations characterize the spatio-temporal evolution of biological systems.*** In contrast to ordinary differential equations, partial differential equations are typically used to study spatial patterns of inherently heterogeneous, regionally varying fields, for example, the flow of blood through the cardiovascular system [Kissas et al., 2018] or the elastodynamic contraction of the heart [Baillargeon et al., 2014]. Unavoidably, these equations are non-linear and highly coupled, and we usually employ computational tools, for example, finite difference or finite element methods, to approximate their solution numerically. Finite element methods have a long history of success at combining ordinary differential equations and partial differential equations to pass knowledge across the scales [De et al., 2014]. They are naturally tailored to represent the small-scale behavior locally through constitutive laws using ordinary differential equations and spatial derivatives and embed this knowledge globally into physics-based conservation laws using partial differential equations. Assuming we know the governing ordinary and partial differential equations, finite element models can predict the behavior of the system from given initial and boundary conditions measured at a few selected points. This approach is incredibly powerful, but requires that we actually know the physics of the system, for example through the underlying kinematic equations, the balance of mass, momentum, or energy. Yet, to close the system of equations, we need constitutive equations that characterize the behavior of the system, which we need to calibrate either with experimental data or with data generated via multiscale modeling.

***Multiscale modeling seeks to predict the behavior of biological, biomedical, and behavioral systems***. Toward this goal, the main objective of multiscale modeling is to identify *causality* and establish causal relations between data. Our experience has taught us that most engineering materials display an elastic, viscoelastic, or elastoplastic constitutive behavior. However, biological and biomedical materials are often more complex, simply because they are alive [Ambrosi et al., 2011]. They continuously interact with and adapt to their environment and dynamically respond to biological, chemical, or mechanical cues [Humphrey et al., 2014]. Unlike classical engineering materials, living matter has amazing abilities to generate force, actively contract, rearrange its architecture, and grow or shrink in size [Goriely, 2017]. To appropriately model these phenomena, we not only have to rethink the underlying kinetics, the balance of mass, and the laws of thermodynamics, but often have to include the biological, chemical, or electrical fields that act as stimuli of this living response [Lorenzo et al., 2016]. This is where multiphysics multiscale modeling becomes important [Southern et al., 2008; Chabiniok et al., 2016]: Multiscale modeling allows us to thoroughly probe biologically relevant phenomena at a smaller scale and seamlessly embed the relevant *mechanisms* at the larger scale to *predict* the dynamics of the overall system [Hunt et al., 2018]. Importantly, rather than making phenomenological assumptions about the behavior at the larger scale, multiscale models postulate that the behavior at the larger scale *emerges* naturally from the collective action at the smaller scales. Yet, this attention to detail comes at a price. While multiscale models can provide unprecedented insight to mechanistic detail, they are not only expensive, but also introduce a large number of unknowns, both in the form of unknown physics and unknown parameters [Raissini & Karniadakis 2018, Raissi et al., 2019]. Fortunately, with the increasing ability to record and store information, we now have access to massive amounts of biological and biomedical data that allow us to systematically discover details about these unknowns.

***Machine learning seeks to infer the dynamics of biological, biomedical, and behavioral systems***. Toward this goal, the main objective of machine learning is to identify *correlations* among big data. The focus in the biology, biomedicine, and behavioral sciences is currently shifting from solving forward problems based on sparse data towards solving inverse problems to explain large data sets [Raissi et al., 2018]. Today, multiscale simulations in the biological, biomedical, and behavioral sciences seek to *infer* the behavior of the system, assuming that we have access to massive amounts of data, while the governing equations and their parameters are not precisely known [Brunton et al., 2016, Raissi et al., 2017d, Wang et al., 2019]. This is where machine learning becomes critical: Machine learning allows us to systematically preprocess massive amounts of data, integrate and analyze it from different input modalities and different levels of fidelity, identify *correlations*, and *infer* the dynamics of the overall system. Similarly, we can use machine learning to quantify the agreement of correlations, for example by comparing computationally simulated and experimentally measured features across multiple scales using Bayesian inference and uncertainty quantification [Sahli et al., 2019].

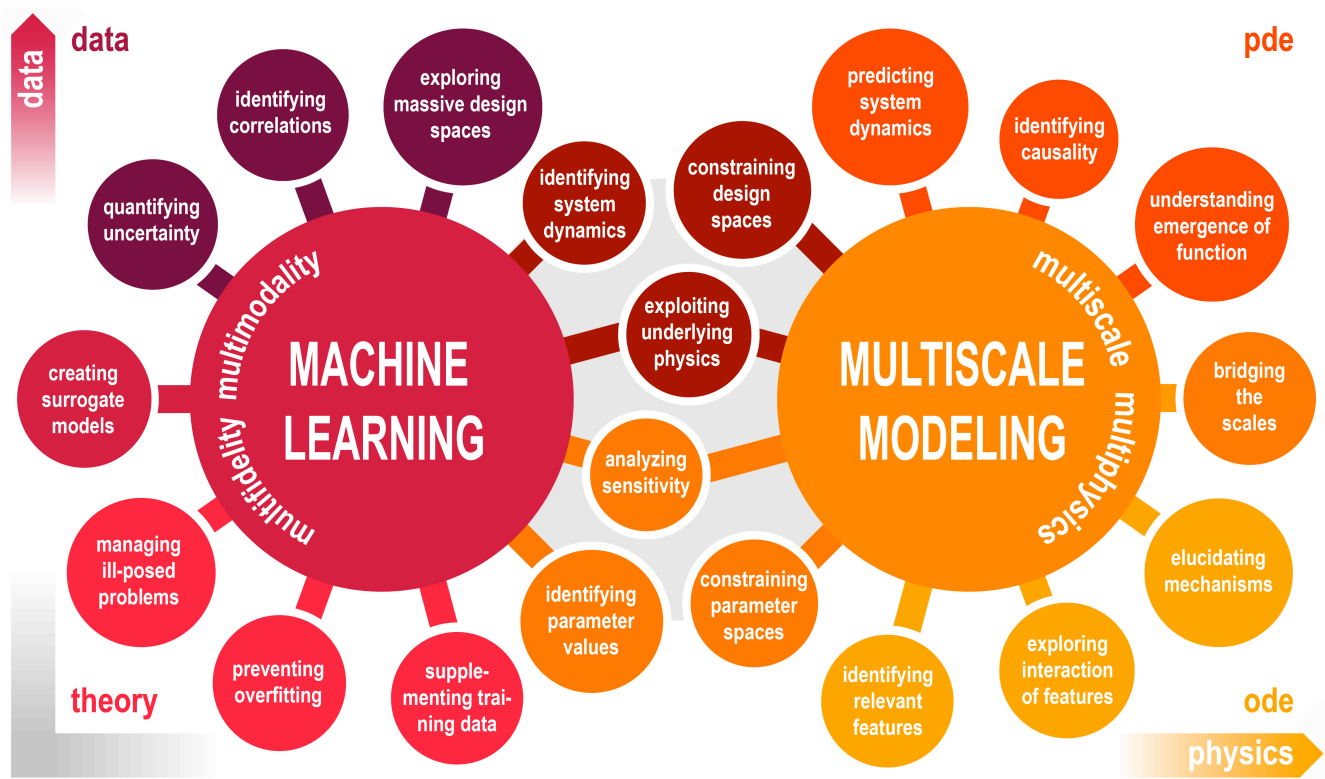

**Figure 1**. **Machine learning and multiscale modeling in the biological, biomedical, and behavioral sciences**. Machine learning and multiscale modeling interact on the parameter level via constraining parameter spaces, identifying parameter values, and analyzing sensitivity and on the system level via exploiting the underlying physics, constraining design spaces, and identifying system dynamics. Machine learning provides the appropriate tools towards supplementing training data, preventing overfitting, managing ill-posed problems, creating surrogate models, and quantifying uncertainty with the ultimate goal being to explore massive design spaces and identify correlations. Multiscale modeling integrates the underlying physics towards identifying relevant features, exploring their interaction, elucidating mechanisms, bridging scales, and understanding the emergence of function with the ultimate goal of predicting system dynamics and identifying causality.

***Machine learning and multiscale modeling mutually complement one another***. Where machine learning reveals correlation, multiscale modeling can probe whether the correlation is causal; where multiscale modeling identifies mechanisms, machine learning, coupled with Bayesian methods, can quantify uncertainty. This natural synergy presents exciting challenges and new opportunities in the biological, biomedical, and behavioral sciences [Lytton et al., 2017]. On a more fundamental level, there is a pressing need to develop the appropriate *theories* to integrate machine learning and multiscale modeling. For example, it seems intuitive to a priori build physics-based knowledge in the form of partial differential equations, boundary conditions, and constraints into a machine learning approach [Raissi et al., 2019]. Especially when the available data are limited, we can increase the robustness of machine learning by including physical constraints such as conservation, symmetry, or invariance. On a more translational level, there is a need to integrate data from different modalities to build predictive simulation tools of biological systems [Perdikaris & Karniadakis, 2016]. For example, it seems reasonable to assume that experimental data from cell and tissue level experiments, animal models, and patient recordings are strongly correlated and obey similar physics-based laws, even if they do not originate from the same system. Naturally, while data and theory go hand in hand, some of the approaches to integrate information are more data driven, seeking to answer questions about the quality of the data, identify missing information, or supplement sparse training data [Tartakovsky et. al, 2018a, 2018b], while some are more theory driven, seeking to answer questions about robustness and efficiency, analyze sensitivity, quantify uncertainty, and choose appropriate learning tools.

Figure 1 illustrates the integration of machine learning and multiscale modeling on the parameter level by constraining their spaces, identifying values, and analyzing their sensitivity, and on the system level by exploiting the underlying physics, constraining design spaces, and identifying system dynamics. Machine learning provides the appropriate tools for supplementing training data, preventing overfitting, managing ill-posed problems, creating surrogate models, and quantifying uncertainty. Multiscale modeling integrates the underlying physics for identifying relevant features, exploring their interaction, elucidating mechanisms, bridging scales, and understanding the emergence of function. We have structured this review around four distinct but overlapping methodological areas: ordinary and partial differential equations, and data and theory driven machine learning. These four themes roughly map into the four corners of the data-physics space, where the amount of available data increases from top to bottom and physical knowledge increases from left to right. For each area, we identify challenges, open questions, and opportunities, and highlight various examples from the life sciences. For convenience, we summarize the most important terms and technologies associated with machine learning with examples from multiscale modeling in Box 1. We envision that our article will spark discussion and inspire scientists in the fields of machine learning and multiscale modeling to join forces towards creating predictive tools to reliably and robustly predict biological, biomedical, and behavioral systems for the benefit of human health.

**CHALLENGES**

A major challenge in the biological, biomedical, and behavioral sciences is to understand systems for which the underlying data are incomplete and the physics are not yet fully understood. In other words, with a complete set of high-resolution data, we could apply machine learning to explore design spaces and identify correlations; with a validated and calibrated set of physics equations and material parameters, we could apply multiscale modeling to predict system dynamics and identify causality. By integrating machine learning and multiscale modeling we can leverage the potential of both, with the ultimate goal of providing quantitative predictive insight into biological systems. Figure 2 illustrates how we could integrate machine learning and multiscale modeling to better understand the cardiac system.

**Active learning** is a supervised learning approach in which the algorithm actively chooses the input training points. When applied to classification, it selects new inputs that lie near the classification boundary or minimize the variance. Example: Classification of arrhythmogenic risk [Sahli Costabal et al., 2019c].

**Bayesian inference** is a method of statistical inference that uses Bayes' theorem to update the probability of a hypothesis as more information becomes available. Examples: Selecting models and identifying parameters of liver [Madireddy et al., 2015], brain [Mihai et al., 2018], and cardiac tissue [Sahli Costabal et al., 2019].

**Classification** is a supervised learning approach in which the algorithm learns from a training set of correctly classified observations and uses this learning to classify new observations, where the output variable is discrete. Examples: Classifying the effects of individual single nucleotide polymorphisms on depression [Athreya et al., 2019]; of ion channel blockage on arrhythmogenic risk in drug development [Sahli Costabal et al., 2019c]; and of chemotherapeutic agents in personalized cancer medicine [Deist et al., 2019].

**Clustering** is an unsupervised learning method that organizes members of a dataset into groups that share common properties. Examples: Clustering the effects of simulated treatments [Lin et al., 2018; Neymotin et al., 2016].

**Convolutional neural networks** are neural network that apply the mathematical operation of convolution, rather than linear transformation, to generate the following output layer. Examples: Predicting mechanical properties using microscale volume elements through deep learning [Yang et al., 2018a], classifying red blood cells in sickle cell anemia [Xu et al., 2017], and inferring the solution of multi-scale partial differential equations [Zhu et al., 2019].

**Deep neural networks** or deep learning are a powerful form of machine learning that uses neural networks with a multiplicity of layers. Examples: biologially-inspired learning, where deep learning aims to replicate mechanisms of neuronal interactions in the brain [Marblestone et al., 2016], predicting the sequence specificities of DNA-and RNA-binding proteins [Alipanahi et. al., 2015].

**Domain randomization** is a technique for randomizing the field of an image so that the true image is also recognized as a realization of this space. Example: Supplementing triaining data [Tremblay et al., 2018].

**Dropout neural networks** are a regularization method for neural networks that avoids overfitting by randomly deleting, or dropping, units along with their connections during training. Examples: Detecting retinal diseases and making diagnosis with various qualities of retinal image data [Rajan et al., 2018]

**Dynamic programming** is a mathematical optimization formalism that enables the simplification of a complicated decision-making problem by recursively breaking it into simpler sub-problems. Example: de novo peptide sequencing via tandem mass spectrometry and dynamic programming [Chen et. al., 2001].

**Evolutionary algorithms** are generic population-based optimization algorithms that adopt mechanisms inspired by biological evolution including reproduction, mutation, recombination, and selection to characterize biological systems. Example: Automatic parameter tuning in multiscale brain modeling [Dura-Bernal et al., 2017].

**Gaussian process regression** is a nonparametric, Bayesian approach to regression to create surrogate models and quantify uncertainty. Examples: Creating surrogate models to characterize the effects of drugs on features of the electrocardiogram [Sahli Costabal et al., 2019a] or of material properties on the stress profiles from reconstructive surgery [Lee et al., 2019a].

**Genetic programming** is a heuristic search technique of evolving programs that starts from a population of random unfit programs and applies operations similar to natural genetic processes to identify a suitable program. Example: Predicting metabolic pathway dynamics from time-series multi-omics data [Costello & Martin, 2018].

**Generative models** are statistical models that aim to capture the joint distribution between a set of observed or latent random variables. Example: Using deep generative models for chemical space exploration and matter engineering [Sanchez-Lengeling et. al., 2019].

**Multi-fidelity modeling** is a supervised learning approach to synergistically combine abundant, inexpensive, low fidelity and sparse, expensive, high fidelity data from experiments and simulations to create efficient and robust surrogate models. Examples: Simulating the mixed convection flow past a cylinder [Perdikaris et al., 2016] and cardiac electrophysiology [Sahli Costabal et al., 2019b]

**Physics-informed neural networks** are neural networks that solve supervised learning tasks while respecting physical constraints. Examples: Diagnosing cardiovascular disorders non-invasively using four-dimensional magnetic resonance images of blood flow and arterial wall displacements [Kissas et al., 2019] and creating computationally efficient surrogates for velocity and pressure fields in intracranial aneurysms [Raissi et al., 2018].

**Recurrent neural networks** are a class of neural networks that incorporate a notion of time by accounting not only for current data, but also for history with tunable extents of memory. Example: Identifying unknown constitutive relations in ordinary differential equation systems [Hagge et al., 2017]

**Reinforcement learning** is a technique that circumvents the notions of supervised and unsupervised learning by exploring and combining decisions and actions in dynamic environments to maximize some notion of cumulative reward. Examples: Understanding common learning modes in biological, cognitive, and artificial systems through the lens of reinforcement learning [Botvinick et. al., 2019, Neftci et. al., 2019].

**Regression** is a supervised learning approach in which the algorithm learns from a training set of correctly identified observations and then uses this learning to evaluate new observations where the output variable is continuous. Example: Exploring the interplay between drug concentration and drug toxicity in cardiac elecrophysiology [Sahli Costabal et al., 2019b].

**Supervised learning** defines the task of learning a function that maps an input to an output based on example input-output pairs. Typical examples include classification and regression tasks.

**System identification** refers to a collection of techniques that identify the governing equations from data on a steady state or dynamical system. Examples: Inferring operators that form ordinary [Mangan et al., 2016] and partial differential equations [Wang et al., 2019].

**Uncertainty quantification** is the science of quantitative characterization and reduction of uncertainties that seeks to determine the likelihood of certain outputs if the inputs are not exactly known. Example: Quantifying the effects of experimental uncertainty in heart failure [Peirlinck et al., 2019] or the effects of estimated material properties on stress profiles in reconstructive surgery [Lee et al. 2018].

**Unsupervised learning** aims at drawing inferences from datasets consisting of input data without labeled responses. The most common types of unsupervised learning techniques include clustering and density estimation used for exploratory data analysis to identify hidden patterns or groupings.

**Box 1.** Terms and technologies associated with machine learning with examples from multiscale modeling in the biological, biomedical, and behavioral sciences.

***Ordinary differential equations encode temporal evolution into machine learning.*** Ordinary differential equations in time are ubiquitous in the biological, biomedical, and behavior sciences. This is largely because it is relatively easy to make observations and acquire data at the molecular, cellular, organ, or population scales without accounting for spatial heterogeneity, which is often more difficult to access. The descriptions typically range from single ordinary differential equations to large systems of ordinary differential equations or stochastic ordinary differential equations. Consequently, the number of parameters is large and can easily reach thousands or more [Yang et al., 2018; Yang & Perdikaris, 2019]. Given adequate data, the challenge begins with identifying the nonlinear, coupled driving terms [Teichert et al., 2019a]. To analyze the data, we can apply formal methods of *system identification*, including *classical regression* and *stepwise regression* [Brunton et al., 2016; Wang et al., 2019]. These approaches are posed as nonlinear optimization problems to determine the set of coefficients by multiplying combinations of algebraic and rate terms that result in the best fit to the observations. Given adequate data, system identification works with notable robustness and can learn a parsimonious set of coefficients, especially when using stepwise regression. In addition to identifying coefficients, the system identification should also address *uncertainty quantification* and account for both measurement errors and model errors. The Bayesian setting provides a formal framework for this purpose [Kennedy & O'Hagan, 2002]. Recent system identification techniques [Brunton et al., 2016; Mangan et al., 2016; Rudy et al., 2017; Quade et al., 2018; Mangan et al., 2019; Champion et al., 2019; Wang et al., 2019] start from a large space of candidate terms in the ordinary differential equations to systematically control and treat model errors. Machine learning can provide a powerful approach to reduce the number of dynamical variables and parameters while maintaining the biological relevance of the model [Brunton et al. 2016, Snowden et al. 2017].

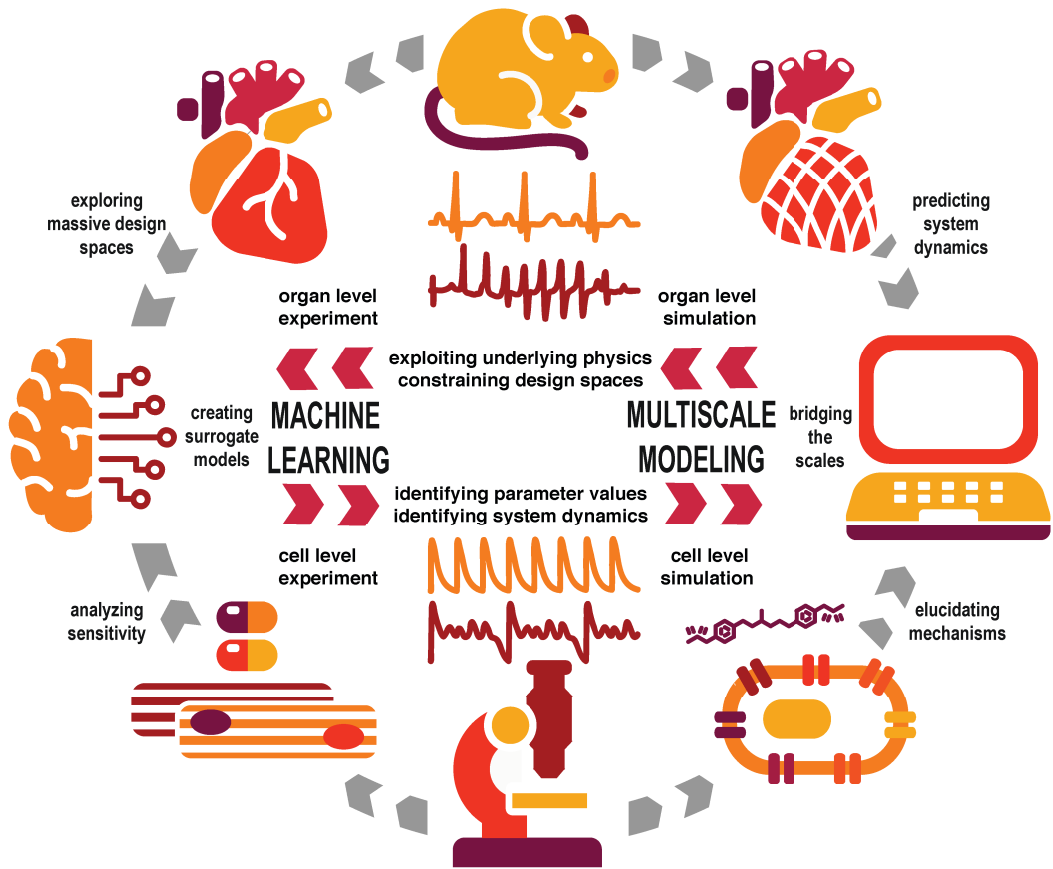

**Figure 2**. **Machine learning and multiscale modeling of the cardiac system**. Multiscale modeling can teach machine learning how to exploit the underlying physics described by, e.g., the ordinary differential equations of cellular electrophysiology and the partial differential equations of electro-mechanical coupling, and constrain the design spaces; machine learning can teach multiscale modeling how to identify parameter values, e.g., the gating variables that govern local ion channel dynamics, and identify system dynamics, e.g., the anisotropic signal propagation that governs global diffusion. This natural synergy presents new challenges and opportunities in the biological, biomedical, and behavioral sciences.

***Partial differential equations encode physics-based knowledge into machine learning***. The interaction between the different scales, from the cell to the tissue and organ levels, is generally complex and involves temporally and spatially varying fields with many unknown parameters [Walpole et al., 2013]. Prior physics-based information in the form of partial differential equations, boundary conditions, and constraints can regularize a machine learning approach in such a way that it can robustly learn from small and noisy data that evolve in time and space. Gaussian processes and neural networks have proven particularly powerful in this regard [E et al., 2017, Raissi et al., 2017a, 2017b]. For Gaussian process regression, the partial differential equation is encoded in an informative function prior [Raissi & Karniadakis, 2018]; for deep neural networks, the partial differential equation induces a new neural network coupled to the standard uninformed data-driven neural network [Raissi et al., 2019], see Figure 3. This coupling of data and partial differential equations into a deep neural network presents itself as an approach to impose physics as a constraint on the expressive power of the latter. New theory driven approaches are required to extend this approach to stochastic partial differential equations using generative adversarial networks, for fractional partial differential equations in systems with memory using high-order discrete formulas, and for coupled systems of partial differential equations in multiscale multiphysics modeling. Multiscale modeling is a critical step, since biological systems typically possess a hierarchy of structure, mechanical properties, and function across the spatial and temporal scales. Over the past decade, modeling multiscale phenomena has been a major point of attention, which has advanced detailed deterministic models and their coupling across scales [De et al., 2014]. Recently, machine learning has permeated into the multiscale modeling of hierarchical engineering materials [Raissi et al., 2017a; Liang et al., 2008; Liu et al., 2019; Le et al., 2015] and into the solution of high-dimensional partial differential equations with deep learning methods [Han et al., 2018; E. et al., 2017, 2018; Raissi et al, 2017c; Teichert et al., 2019; Teichert & Garikipati, 2019; Topol, 2019b]. Uncertainty quantification in material properties is also gaining relevance [Hurtado et al., 2017], with examples of Bayesian model selection to calibrate strain energy functions [Mihai et al., 2018; Madireddy et al., 2015] and uncertainty propagation with Gaussian processes of nonlinear mechanical systems [Lee et al., 2019; Lee et al., 2018; Sahli Costabal et al., 2019]. These trends for non-biological systems point towards immediate opportunities for integrating machine learning and multiscale modeling in the biological, biomedical, and behavioral sciences and opens new perspectives that are unique to the living nature of biological systems.

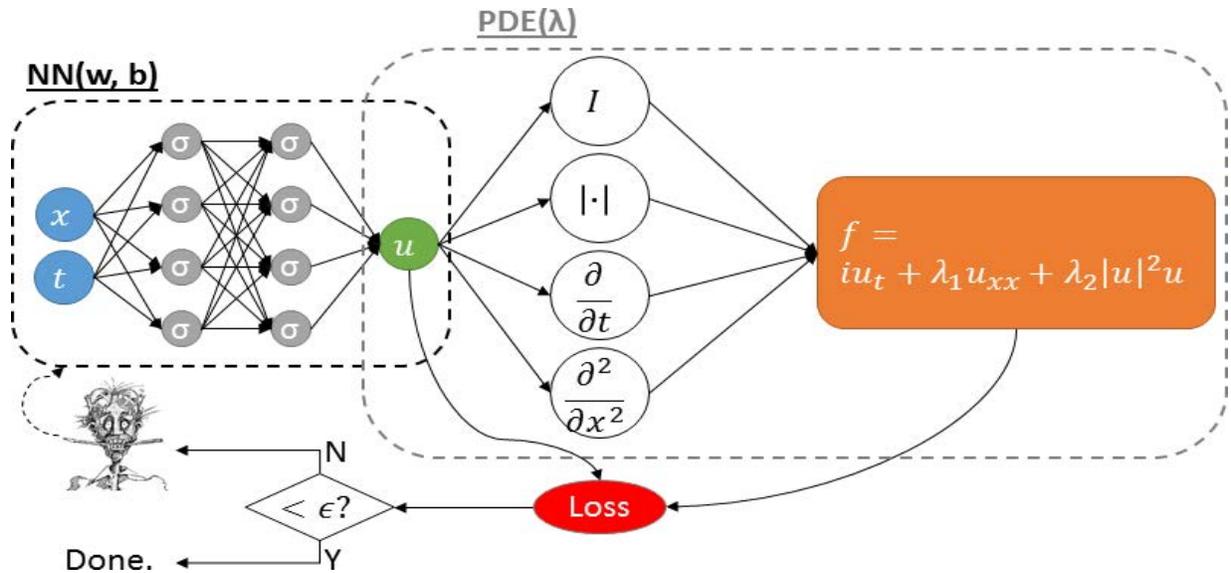

**Figure 3**. **Partial differential equations encode physics-based knowledge into machine learning.** Physics imposed on neural networks. The neural network on the left, as yet unconstrained by physics, represents the solution u(x,t) of the partial differential equation; the neural network on the right describes the residual f(x,t) of the partial differential equation. The example illustrates a one-dimensional version of the Schrödinger equation with unknown parameters $\lambda_1$ and $\lambda_2$ to be learned. In addition to unknown parameters, we can learn missing functional terms in the partial differential equation. Currently, this optimization is done empirically based on trial and error by a human-in-the-loop. Here, the u-architecture is a fully-connected neural network, while the f-architecture is dictated by the partial differential equation and is, in general, not possible to visualize explicitly. Its depth is proportional to the highest derivative in the partial differential equation times the depth of the uninformed u neural network.

*Data-driven machine learning seeks correlations in big data.* Machine learning can be regarded as an extension of classical statistical modeling that can digest massive amounts of data to *identify high-order correlations* and generate predictions. This is not only important in view of the rapid developments of ultra-high resolution measurement techniques [van den Bedem & Fraser, 2015], including cryo-EM, high-resolution imaging flow cytometry, or four-dimensional-flow magnetic resonance imaging, but also when analyzing large-scale health data from wearable and smartphone apps [Althoff et al., 2017; Hicks et al., 2019]. Machine learning can play a central role in helping us mine these data more effectively and bring experiment, modeling, and computation closer together [Duraisamy et al., 2019]. We can use machine learning as a tool in developing artificial intelligence applications to solve complex biological, biomedical, or behavioral systems [Topol, 2019a]. Figure 4 illustrates a framework for integrating machine learning and multiscale modeling with a view towards data-driven approaches. Most data-driven machine learning techniques seek *correlation* rather than *causality*. Some machine learning techniques, e.g., Granger causality [Tank et al., 2018] or dynamic causal modeling [Friston et al., 2003], do seek causality, but without mechanisms. In contrast to machine learning, multiscale modeling seeks to provide not only correlation or causality but also the underlying *mechanism* [Hunt et al., 2018]. This suggests that machine learning and multiscale modeling can effectively complement one another when analyzing big data: Where machine learning reveals a correlation, multiscale modeling can probe whether this correlation is causal, and can unpack cause into mechanisms or mechanistic chains at lower scales [Lytton et al., 2017]. This unpacking is particularly important in personalized medicine where each patient's disease process is a unique variant, traditionally lumped into large populations by evidence based medicine, whether through the use of statistics, machine learning, or artificial intelligence. Multiscale models can split the variegated patient population apart by identifying mechanistic variants based on differences in genome of the patient, as well as genomes of invasive organisms or tumor cells, or immunological history. This is an important step towards creating a digital twin, a multiscale model of an organ system or a disease process, where we can develop therapies without risk to the patient. As multiscale modeling attempts to leverage the vast volume of experimental data to gain understanding, machine learning will provide invaluable tools to preprocess these data, automate the construction of models, and analyze the similarly vast output data generated by multiscale modeling [Dura-Bernal et al., 2019; Vu et al., 2018].

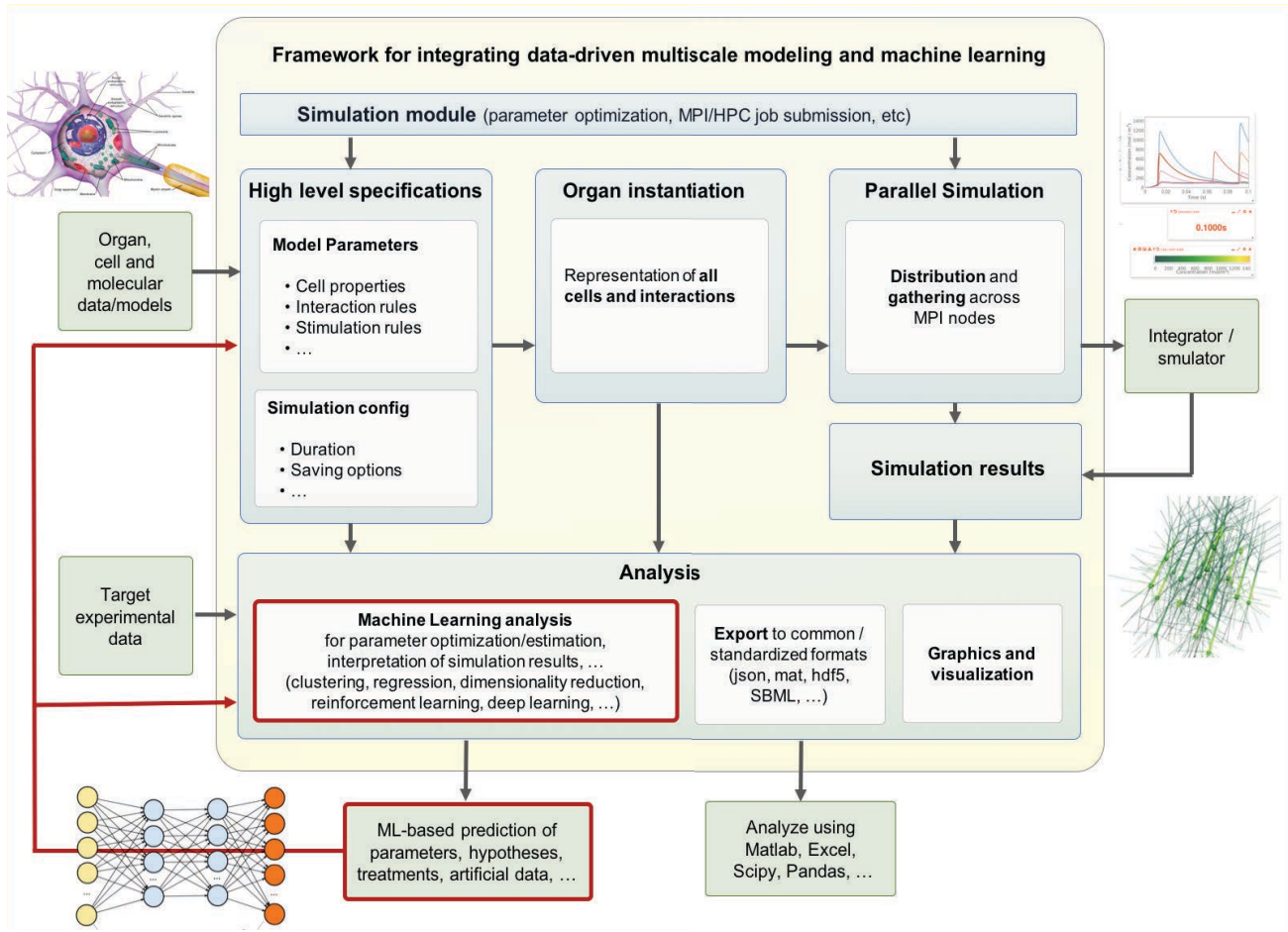

**Figure 4. Data-driven machine learning seeks correlations in big data.** This general framework integrates data-driven multiscale modeling and machine learning by performing organ, cellular, or molecular level simulations and systematically comparing the simulation results against experimental target data using machine learning analysis including clustering, regression, dimensionality reduction, reinforcement learning, and deep learning with the objectives to identify parameters, generate new hypotheses, or optimize treatment.

***Theory-driven machine learning seeks causality by integrating physics and big data***. The basic idea of theory-driven machine learning is, given a physics-based ordinary or partial differential equation, how can we leverage structured physical laws and mechanistic models as informative prior information in a machine learning pipeline towards advancing modeling capabilities and expediting multiscale simulations? Figure 5 illustrates the integration of theory-driven machine learning and multiscale modeling to accelerate model- and data-driven discovery. Historically, we have solved this problem using dynamic programing and variational methods. Both are extremely powerful when we know the physics of the problem and can constrain the parameters space to reproduce experimental observations. However, when the underlying physics are unknown, or there is uncertainty about their form, we can adapt machine learning techniques that learn the underlying system dynamics. Theory-driven machine learning allows us to seamlessly integrate physics-based models at multiple temporal and spatial scales. For example, *multifidelity techniques* can combine coarse measurements and reduced order models to significantly accelerate the prediction of expensive experiments and large-scale computations [Perdikaris & Karniadakis, 2016; Perdikaris et al., 2016]. In drug development, for example, we can leverage theory-driven machine learning techniques to integrate information across ten orders of magnitude in space and time towards developing interpretable classifiers to characterize the pro-arrhythmic potential of drugs [Sahli Costabal et al., 2019a]. Specifically, we can employ *Gaussian process regression* to effectively explore the interplay between drug concentration and drug toxicity using coarse, low-cost models, anchored by a few, judiciously selected, high-resolution simulations [Sahli Costabal et al., 2019b]. Theory-driven machine learning techniques can also leverage probabilistic formulations to inform the judicious acquisition of new data and actively expedite tasks such as *exploring massive design spaces* or *identifying system dynamics*. For example, we could devise an effective data acquisition policy for choosing the most informative mesoscopic simulations that need to be performed to recover detailed constitutive laws as appropriate closures for macroscopic models of complex fluids [Zhao et al., 2018]. More recently, efforts have been made to directly bake-in theory into machine learning practice. This enables the construction of predictive models that adhere to the underlying physical principles, including conservation, symmetry, or invariance, while remaining robust even when the observed data are very limited. For example, a recent model only utilized conservation laws of reaction to model the metabolism of a cell. While the exact functional forms of the rate laws was unknown, the equations were solved using machine learning [Costello & Martin, 2018]. An intriguing implication is related to their ability to leverage auxiliary observations to infer quantities of interest that are difficult to measure in practice [Raissi et al., 2019]. Another example includes the use of neural networks constrained by physics to infer the arterial blood pressure directly and non-invasively from four-dimensional magnetic resonance images of blood velocities and arterial wall displacements by leveraging the known dynamic correlations induced by first principles in fluid and solid mechanics [Kissas et al., 2019]. In personalized medicine, we can use theory-driven machine learning to classify patients into specific treatment regimens. While this is typically done by genome profiling alone, models that supplement the training data using simulations based on biological or physical principles can have greater classification power than models built on observed data alone. For the examples of radiation impact on cells and Boolean cancer modeling, a recent study has shown that, for small training datasets, simulation-based kernel methods that use approximate simulations to build a kernel improve the downstream machine learning performance and are superior over standard no-prior-knowledge machine learning techniques [Deist et al., 2019].

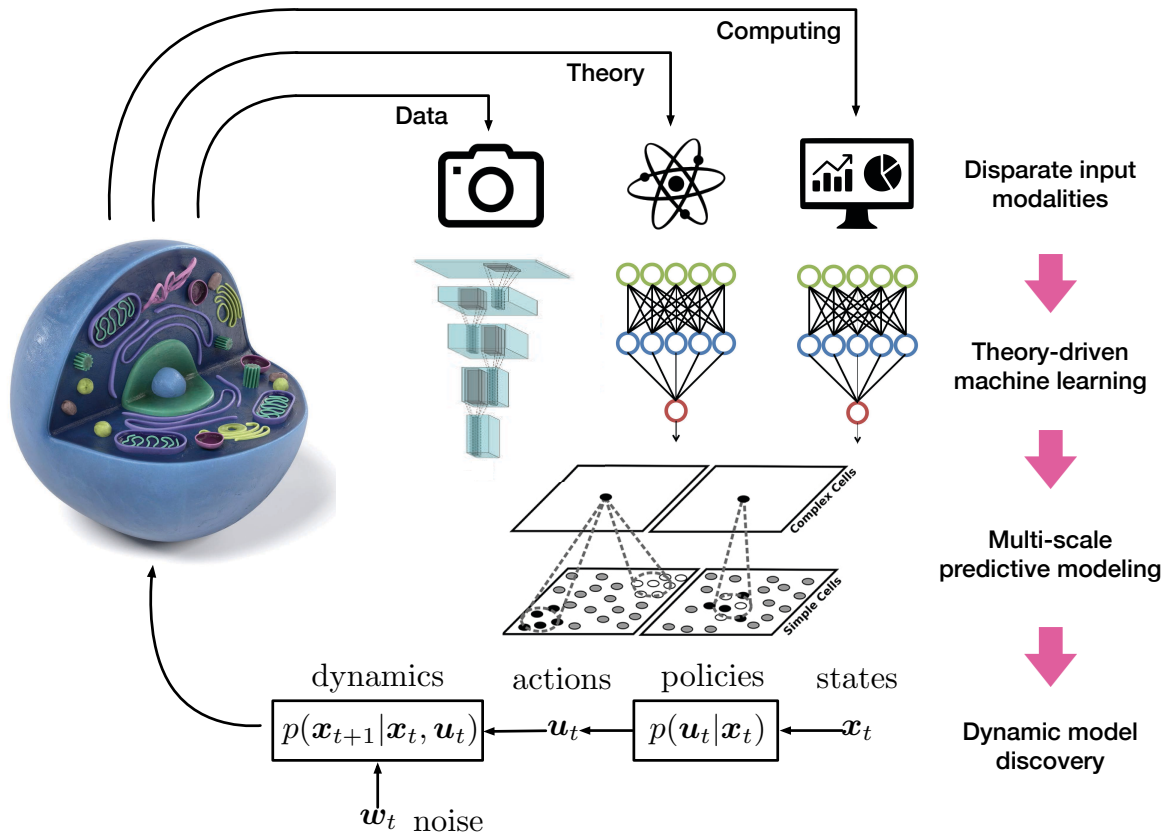

**Figure 5**. **Theory-driven machine learning seeks causality by integrates prior knowledge and big data.** Accelerating model- and data-driven discovery by integrating theory-driven machine learning and multiscale modeling. Theory-driven machine learning can yield data-efficient workflows for predictive modeling by synthesizing prior knowledge and multimodality data at different scales. Probabilistic formulations can also enable the quantification of predictive uncertainty and guide the judicious acquisition of new data in a dynamic model-refinement setting.

**OPEN QUESTIONS AND OPPORTUNITIES**

Numerous open questions and opportunities emerge from integrating machine learning and multiscale modeling in the biological, biomedical, and behavioral sciences. We address some of the most urgent ones below.

***Managing ill-posed problems.*** Can we solve ill-posed inverse problems that arise during parameter or system identification? Unfortunately, many of the inverse problems for biological systems are ill posed. Mathematically speaking, they constitute boundary value problems with unknown boundary values. Classical mathematical approaches are not suitable in these cases. Methods for backward uncertainty quantification could potentially deal with the uncertainty involved in inverse problems, but these methods are difficult to scale to realistic settings. In view of the high dimensional input space and the inherent uncertainty of biological systems, inverse problems will always be challenging. For example, it is difficult to determine if there are multiple solutions or no solutions at all, or to quantify the confidence in the prediction of an inverse problem with high-dimensional input data. Does the inherent regularization in the loss function of neural networks allow us to deal with ill-posed inverse partial differential equations without boundary or initial conditions and discover hidden states?

***Identifying missing information***. Are the parameters of the proposed model sufficient to provide a basic set to produce higher-scale system dynamics? Multiscale simulations and generative networks can be set up to work in parallel, alongside the experiment, to provide an independent confirmation of parameter sensitivity. For example, circadian rhythm generators provide relatively simple dynamics but have very complex dependence on numerous underlying parameters, which multiscale modeling can reveal. An open opportunity exists to use generative models to identify both the underlying low dimensionality of the dynamics and the high dimensionality associated with parameter variation. Inadequate multiscale models could then be identified with failure of generative model predictions.

***Creating surrogate models***. Can we use generative adversarial networks to create new test data sets for multiscale models? Conversely, can we use multiscale modeling to provide training or test instances to create new surrogate models using deep learning? By using deep learning networks, we could provide answers more quickly than by using complex and sophisticated multiscale models. This could, for example, have significant applications in predicting pharmaceutical efficacy for patients with particular genetic inheritance in personalized medicine.

***Discretizing space and time***. Can we remove or automate the tyranny of grid generation in conventional methods? Discretization of complex and moving three-dimensional domains remains a time- and labor-intense challenge. It generally requires specific expertise and many hours of dedicated labor, and has to be re-done for every individual model. This becomes particularly relevant when creating personalized models with complex geometries at multiple spatial and temporal scales. While many efforts in machine learning are devoted to solving partial differential equations in a given domain, new opportunities arise for machine learning when dealing directly with the creation of the discrete problem. This includes automatic mesh generation, meshless interpolation, and parameterization of the domain itself as one of the inputs for the machine learning algorithm. Neural networks constrained by physics remove the notion of a mesh, but retain the more fundamental concept of basis functions: They impose the conservation laws of mass, momentum, and energy at, e.g., collocation points that, while neither connected through a regular lattice nor through an unstructured grid, serve to determine the parameters that define the basis functions.

*Bridging the scales*. Can machine learning provide scale bridging in cases where a relatively clean separation of scales is possible? For example, in cancer, machine learning could be used to explore responses of both immune cells and tumor cells based on single-cell data. This example points towards opportunities to build a multiscale model on the families of solutions to codify the evolution of the tumor at the organ or metastasis scales.

*Supplementing training data.* Can we use simulated data to supplement training data? Supervised learning, as used in deep networks, is a powerful technique, but requires large amounts of training data. Recent studies have shown that, in the area of object detection in image analysis, simulation augmented by domain randomization can be used successfully as a supplement to existing training data. In areas where multiscale models are well-developed, simulation across vast areas of parameter can, for example, supplement existing training data for nonlinear diffusion models to provide physics-informed machine learning. Similarly, multiscale models can be used in biological, biomedical, and behavioral systems to augment insufficient experimental or clinical data sets.

*Quantifying uncertainty*. Can theory-driven machine learning approaches enable the reliable characterization of predictive uncertainty and pinpoint its sources? Uncertainty quantification is the backbone of decision-making. This has many practical applications such as decision-making in the clinic, the robust design of synthetic biology pathways, drug target identification and drug risk assessment. There are also opportunities to use quantification to guide the informed, targeted acquisition of new data.

*Exploring massive design spaces*. Can theory-driven machine learning approaches uncover meaningful and compact representations for complex inter-connected processes, and, subsequently, enable the cost-effective exploration of vast combinatorial spaces? While this is already pretty common in the design of bio-molecules with target properties in drug development, there many other applications in biology and biomedicine that could benefit from these technologies.

*Elucidating mechanisms*. Can theory-driven machine learning approaches enable the discovery of interpretable models that can not only explain data, but also elucidate mechanisms, distill causality, and help us probe interventions and counterfactuals in complex multiscale systems? For instance, causal inference generally uses various statistical measures such as partial correlation to infer causal influence. If instead, the appropriate statistical measure were known from the underlying physics, would the causal inference be more accurate or interpretable as a mechanism?

*Understanding emergence of function*. Can theory-driven machine learning, combined with sparse and indirect measurements, produce a mechanistic understanding of the emergence of biological function? Understanding the emergence of function is of critical importance in biology and medicine, environmental studies, biotechnology, and other biological sciences. The study of emergence critically relies on our ability to model collective action on a lower scale to predict how the phenomena on the higher scale emerges from this collective action.

*Harnessing biologically-inspired learning.* Can we harness biological learning to design more efficient algorithms and architectures? Artificial intelligence through deep learning is an exciting recent development that has seen remarkable success in solving problems, which are difficult for humans. Typical examples include chess and Go, as well as the classical problem of image recognition, that, although superficially easy, engages broad areas of the brain. By contrast, activities that neuronal networks are particularly good at remain beyond the reach of these techniques, for example, the control systems of a mosquito engaged in evasion and targeting are remarkable considering the small neuronal network involved. This limitation provides opportunities for more detailed brain models to assist in developing new architectures and new learning algorithms.

*Preventing overfitting*. Can we use prior physics based knowledge to avoid overfitting or non-physical predictions? How can we calibrate and validate the proposed models without overfitting? How can we apply cross-validation to simulated data, especially when the simulations may contain long-time correlations? From a conceptual point of view, this is a problem of supplementing the set of known physics-based equations with constitutive equations, an approach, which has long been used in traditional engineering disciplines. While data-driven methods can provide solutions that are not constrained by preconceived notions or models, their predictions should not violate the fundamental laws of physics. Sometimes it is difficult to determine whether the model predictions obey these fundamental laws, especially when the functional form of the model cannot be determined explicitly. This makes it difficult to know whether the analysis predicts the correct answer for the right reasons. There are well-known examples of deep learning neural networks that appear to be highly accurate, but make highly inaccurate predictions when faced with data outside their training regime, and others that make highly inaccurate predictions based on seemingly minor changes to the target data. To address this limitation, there are numerous opportunities to combine machine learning and multiscale modeling towards a priori satisfying the fundamental laws of physics, and, at the same time, preventing overfitting of the data.

*Minimizing data bias*. Can an arrhythmia patient trust a neural net controller embedded in a pacemaker that was trained under different environmental conditions than the ones during his own use? Training data come at various scales and different levels of fidelity. Data are typically generated by existing models, experimental assays, historical data, and other surveys, all of which come with their own inductive biases. Machine learning algorithms can only be as good as the data they have seen. This implies that proper care needs to be taken to safe-guard against biased data-sets. New theory-driven approaches could provide a rigorous foundation to estimate the range of validity, quantify the uncertainty, and characterize the level of confidence of machine learning based approaches.

*Increasing rigor and reproducibility*. Can we establish rigorous validation tests and guidelines to thoroughly test the predictive power of models built with machine learning algorithms? The use of open source codes and data sharing by the machine learning community is a positive step, but more benchmarks and guidelines are needed for neural networks constrained by physics. Reproducibility has to be quantified in terms of statistical metrics, as many optimization methods are stochastic in nature and may lead to different results. In addition to memory, the 32-bit limitation of current GPU systems is particularly troubling for modeling biological systems where steep gradients and very fast multirate dynamics may require 64-bit arithmetic, which, in turn, may require ten times more computational time with the current technologies.

## CONCLUSIONS

Machine learning and multiscale modeling naturally complement and mutually benefit from one another. Machine learning can explore massive design spaces to identify correlations and multiscale modeling can predict system dynamics to identify causality. Recent trends suggest that integrating machine learning and multiscale modeling could become key to better understand biological, biomedical, and behavioral systems. Along those lines, we have identified five major challenges in moving the field forward.

*The first challenge is to create robust predictive mechanistic models when dealing with sparse data*. The lack of sufficient data is a common problem in modeling biological, biomedical, and behavioral systems. For example, it can result from an inadequate experimental resolution or an incomplete medical history. A critical first step is to systematically *identify the missing information*. Experimentally, this can guide the judicious acquisition of new data or even the design of new experiments to complement the knowledge base. Computationally, this can motivate *supplementing the available training data* by performing computational simulations. Ultimately, the challenge is to maximize information gain and optimize efficiency by combining low- and high-resolution data and integrating data from different sources, which, in machine learning terms, introduces a *multifidelity*, *multimodality* approach.

*The second challenge is to manage ill-posed problems*. Unfortunately, ill-posed problems are relatively common in the biological, biomedical, and behavioral sciences and can result from inverse modeling, for example, when *identifying parameter values* or *identifying system dynamics*. A potential solution is to combine deterministic and stochastic models. Coupling the deterministic equations of classical physics—the balance of mass, momentum, and energy—with the stochastic equations of living systems—cell-signaling networks or reaction-diffusion equations—could help guide the design of computational models for problems that are otherwise ill-posed. Along those lines, physics-informed neural networks and physics-informed deep learning are promising approaches that inherently use *constrained parameter spaces* and *constrained design spaces* to *manage ill-posed problems*. Beyond improving and combining existing techniques, we could even think of developing entirely novel architectures and new algorithms to understand ill-posed biological problems inspired by biological learning.

***The third challenge is to efficiently explore massive design spaces to identify correlations***. With the rapid developments in gene sequencing and wearable electronics, the personalized biomedical data has become as accessible and inexpensive as never before. However, efficiently analyzing big data sets within massive design spaces remains a logistic and computational challenge. Multiscale modeling allows us to integrate physics-based knowledge to *bridge the scales* and efficiently pass information across temporal and spatial scales. Machine learning can utilize these insights for efficient model reduction towards *creating surrogate models* that drastically reduce the underlying parameter space. Ultimately, the efficient analytics of big data, ideally in real time, is a challenging step towards bringing artificial intelligence solutions into the clinic.

***The fourth challenge is to robustly predict system dynamics to identify causality***. Indeed, this is the actual driving force behind integrating machine learning and multiscale modeling for biological, biomedical, and behavioral systems. Can we eventually utilize our models to *identify relevant biological features* and *explore their interaction* in real time? A very practical example of immediate translational value is whether we can identify disease progression biomarkers and *elucidate mechanisms* from massive data sets, for example, early biomarkers of neurodegenerative disease, by *exploiting the fundamental laws of physics*. On a more abstract level, the ultimate challenge is to advance data- and theory-driven approaches to create a mechanistic *understanding of the emergence of biological function* to explain phenomena at higher scale as a result of the collective action on lower scales.

***The fifth challenge is to know the limitations of machine learning and multiscale modeling***. Important steps in this direction are *analyzing sensitivity* and *quantifying of uncertainty*. While machine learning tools are increasingly used to perform sensitivity analysis and uncertainty quantification for biological systems, they are at a high risk of overfitting and generating non-physical predictions. Ultimately, our approaches can only be as good as the underlying models and the data they have been trained on, and we have to be aware of model limitations and data bias. *Preventing overfitting*, *minimizing data bias*, and *increasing rigor and reproducibility* have been and will always remain the major challenges in creating predictive models for biological, biomedical, and behavioral systems.

**Acknowledgments.** This work was inspired by the 2019 Symposium on Integrating Machine Learning with Multiscale Modeling for Biological, Biomedical, and Behavioral Systems (ML-MSM) as part of the Interagency Modeling and Analysis Group (IMAG), and is endorsed by the Multiscale Modeling (MSM) Consortium, by the U.S. Association for Computational Mechanics (USACM) Technical Trust Area Biological Systems, and by the U.S. National Committee on Biomechanics (USNCB). The authors acknowledge the stimulating discussions within these communities.

**Author contributions.** MA, ABT, WC, SD, SDB, KG, GK, WWL, PP, LP, and EK discussed and wrote this manuscript.

**Competing interests.** The authors have no competing interest.

**REFERENCES.**

[1] Alipanahi, B., Delong, A., Weirauch, M. T., & Frey, B. J. Predicting the sequence specificities of DNA-and RNA-binding proteins by deep learning. Nature Biotechnology, 33(8), 831 (2015).

[2] Althoff, T., Hicks, J. L., King, A. C., Delp, S. L., Leskovec, J. Large-scale physical activity data reveal worldwide activity inequality. Nature 547, 336-339 (2017).

[3] Ambrosi, D., Ateshian, G.A., Arruda, E.M., Cowin, S.C., Dumais, J., Goriely, A., Holzapfel, G.A., Humphrey, J.D., Kemkemer, R., Kuhl, E., Olberding, J.E., Taber, L.A. & Garikipati, K. Perspectives on biological growth and remodeling. Journal of the Mechanics and Physics of Solids 59, 863-883 (2011).

[4] Athreya, A. P., Neavin, D., Carrillo-Roa, T., Skime, M., Biernacka, J., Frye, M. A., Rush, A. J., Wang, L., Binder, E. B., Iyer, R. K., Weinshilboum, R. M., & Bobo, W. V. Pharmacogenomics-driven prediction of antidepressant treatment outcomes: A machine learning approach with multi-trial replication. Clinical Pharmacology and Therapeutics. doi:10.1002/cpt.1482 (2019).

[5] Baillargeon, B., Rebelo, N., Fox, D.D., Taylor, R.L. & Kuhl E. The Living Heart Project: A robust and integrative simulator for human heart function. European Journal of Mechanics A/Solids 48, 38-47 (2014).

[6] Botvinick, M., Ritter, S., Wang, J. X., Kurth-Nelson, Z., Blundell, C., & Hassabis, D. Reinforcement learning, fast and slow. Trends in Cognitive Sciences (2019).

[7] Brunton, S.L., Proctor, J.L. & Kutz, J.N. Discovering governing equations from data by sparse identification of nonlinear dynamical systems. Proceedings of the National Academy of Sciences 113, 3932-3937 (2016).

[8]     Bruynseels, K., Santoni de Sio, F. & van den Hoven, J. Digital Twins in health care: Ethical implications of an emerging engineering paradigm. Frontiers in Genetics 9,31 (2018).
[9]     Chen, T., Kao, M. Y., Tepel, M., Rush, J., & Church, G. M. A dynamic programming approach to de novo peptide sequencing via tandem mass spectrometry. Journal of Computational Biology, 8, 325-337 (2001).
[10]    Chabiniok, R., Wang, V., Hadjicharalambous, M., Asner, L., Lee, J., Sermesant, M., Kuhl, E., Young, A., Moireau, P., Nash, M., Chapelle, D. & Nordsletten, D. A. Multiphysics and multiscale modeling, data-model fusion and integration of organ physiology in the clinic: ventricular cardiac mechanics. Interface Focus 6, 20150083 (2016).
[11]    Champion, K. P., Brunton, S. L. & Kutz, J. N. Discovery of nonlinear multiscale systems: Sampling strategies and embeddings. SIAM Journal of Applied Dynamical Systems 18 (2019).
[12]    Costello, Z. & Martin, H. G. A machine learning approach to predict metabolic pathway dynamics from time-series multiomics data. NPJ Systems Biology Applications 4, 19 (2018).
[13]    Cuperlovic-Culf, M. Machine learning methods for analysis of metabolic data and metabolic pathway modeling. Metabolites 8, 4 (2018).
[14]    De, S., Wongmuk, H. & Kuhl, E. editors. Multiscale Modeling in Biomechanics and Mechanobiology. Springer; 2014.
[15]    Deist, T. M., Patti, A., Wang, Z., Krane, D., Sorenson, T. & Craft, D. Simulation assisted machine learning. Bioinformatics, doi: 10.1093/bioinformatics/btz199 (2019).
[16]    Dura-Bernal, S., Neymotin, S. A., Kerr, C. C., Sivagnanam, S., Majumdar, A., Francis, J. T., & Lytton, W. W. Evolutionary algorithm optimization of biological learning parameters in a biomimetic neuroprosthesis. IBM Journal of Research and Development 61(6), 1-14 (2017).
[17]    Dura-Bernal, S., Suter, B. A., Gleeson, P., Cantarelli, M., Quintana, A., Rodriguez, F., Kedziora, D. J., Chadderdon, G. L., Kerr, C. C., Neymotin, S. A., McDougal, R. A., Hines, M., Sheperd, G. M. & Lytton, W. W. NetPyNE, a tool for data-driven multiscale modeling of brain circuits. eLife, 8. https://doi.org/10.7554/eLife.44494 (2019).
[18]    Duraisamy, K., Iaccarino, G. & Xiao, H. Turbulence modeling in the age of data. Annual Review of Fluid Mechanics 51: 1-23 (2019).
[19]    E, W., Han, J. & Jentzen, A. Deep learning-based numerical methods for high-dimensional parabolic partial differential equations and backward stochastic differential equations. Communications in Mathematics and Statistics 5(4), 349-380 (2017).
[20]    E, W., & Yu, B. The deep Ritz method: A deep learning-based numerical algorithm for solving variational problems. Communications in Mathematics and Statistics 6(1), 1-12 (2018).
[21]    Friston, K. J., Harrison, L., & Penny, W. Dynamic causal modelling. NeuroImage 19(4), 1273–1302 (2019).
[22]    Goriely, A. The Mathematics and Mechanics of Biological Growth. Springer. 2017.
[23]    Han, J., Jentzen, A. & E, W. Solving high-dimensional partial differential equations using deep learning. Proceedings of the National Academy of Sciences 115(34) 8505-8510 (2018).
[24]    Hicks, J. L., Althoff, T., Sosic, R., Kuhar, P., Bostjancic, B., King, A. C., Leskovec, J. & Delp, S. L. Best practices for analyzing large-scale health data from wearables and smartphone apps. npj Digital Medicine 2, 45 (2019).
[25]    Humphrey, J. D., Dufresne, E. R. & Schwartz, M. A. Mechanotransduction and extracellular matrix homeostasis. Nature Reviews Molecular Cell Biology 15(12), 802-812 (2014).
[26]    Hunt, C. A., Erdemir, A., Lytton, W. W., Mac Gabhann, F., Sander, E. A., Transtrum, M. K., & Mulugeta, L. The spectrum of mechanism-oriented models and methods for explanations of biological phenomena. Processes 6(5), 56 (2018).
[27]    Hurtado, D. E., Castro, S., Madrid, P. Uncerainty quantification of two models of cardiac electromechanics. International Journal for Numerical Methods in Biomedical Engineering 33, e2894 (2017).
[28]    Kennedy, M. & O'Hagan, A. (2001). Bayesian calibration of computer models (with discussion). Journal of the Royal Statistical Society, Series B. 63, 425-464.
[29]    Kissas, G., Yang, Y., Hwuang, E., Witschey, W. R., Detre, J. A. & Perdikaris, P. Machine learning in cardiovascular flows modeling: Predicting pulse wave propagation from non-invasive clinical measurements using physics-informed deep learning. arXiv preprint arXiv:1905.04817 (2019).
[30]    Le, B. A., Yvonnet, J. & He, Q. C. Computational homogenization of nonlinear elastic materials using neural networks. International Journal for Numerical Methods in Engineering 104, 1061–1084 (2015).
[31]    Lee, T., Turin, S. Y., Gosain, A. K., Bilionis, I. & Buganza Tepole, A. Propagation of material behavior uncertainty in a nonlinear finite element model of reconstructive surgery. Biomechanics and Modeling in Mechanobiology 17(6), 1857-18731 (2018).
[32]    Lee, T., Gosain, A.K., Bilionis, I. & Buganza Tepole, A. Predicting the effect of aging and defect size on the stress profiles of skin from advancement, rotation and transposition flap surgeries. Journal of the Mechanics and Physics of Solids 125, 572–590 (2019).
[33]    Liang, G. & Chandrashekhara, K. Neural network based constitutive model for elastomeric foams. Engineering Structures 30, 2002–2011 (2008).
[34]    Liu, Y. Zhang, L., Yang, Y., Zhou, L., Ren, L., Liu, R., Pang, Z., Deen, M.J. A novel cloud-based framework for the elderly healthcare services using Digital Twin. IEEE Access 7, 49088-49101 (2019).

[35] Lorenzo, G., Scott, M. A., Tew, K., Hughes, T. J. R., Zhang, Y. J., Liu, L., Vilanova, G., Gomez, H. Tissue-scale, personalized modeling and simulation of prostate cancer growth. Proceedings of the National Academy of Sciences 113(48), E7663-E7671 (2016).

[36] Lytton, W. W., Arle, J., Bobashev, G., Ji, S., Klassen, T. L., Marmarelis, V. Z., Schwaber, J., Sherif, M. A. & Sanger, T. D. Multiscale modeling in the clinic: diseases of the brain and nervous system. Brain Informatics 4(4), 219–230 (2017).

[37] Madireddy, S., Sista, B. & Vemaganti, K. A Bayesian approach to selecting hyperelastic constitutive models of soft tissue. Computer Methods in Applied Mechanics and Engineering 291, 102–122 (2015).

[38] Madni, A.M., Madni, C.C. & Lucerno, S.D. Leveraging Digital Twin technology in model-based systems enginereering. Systems 7:1-13 (2019)

[39] Mangan, N. M., Brunton, S. L., Proctor, J. L. & Kutz, J. N. Inferring biological networks by sparse identi_cation of nonlinear dynamics. IEEE Transactions on Molecular, Biological and Multi-Scale Communications 2, 52-63 (2016).

[40] Mangan, N. M., Askham, T., Brunton, S. L., Kutz, N. N. & Proctor, J. L. Model selection for hybrid dynamical systems via sparse regression. Proceedings of the Royal Society A: Mathematical, Physical and Engineering Sciences 475, 20180534 (2019).

[41] Marblestone, A.H., Wayne, G. and Kording, K.P. Toward an integration of deep learning and neuroscience. Front. Comput. Neurosci., doi:10.3389/fncom.2016.00094 (2016).

[42] Mihai, L. A., Woolley, T.E. & Goriely, A. Stochastic isotropic hyperelastic materials: Constitutive calibration and model selection. Proceedings of the Royal Society A / Mathematical, Physical, and Engineering Sciences 474, 0858 (2018).

[43] Nazari, F., Pearson, A. T., Nor, J. E. & Jackson, T. L. A mathematical model for IL-6-mediated, stem cell driven tumor growth and targeted treatment. PLOS Computational Biology, 14(1):e1005920 (2018).

[44] Neftci, E. O., & Averbeck, B. B. Reinforcement learning in artificial and biological systems. Nature Machine Intelligence 1, 133-143 (2019).

[45] Peirlinck, M., Sahli Costabal, F., Sack, K.L., Choy, J.S., Kassab, G.S., Guccione, J.M., De Beule, M., Segers, P., Kuhl, E. Using machine learning to characterize heart failure across the scales. Biomech Model Mechanobio. doi:10.1007/s10237-019-01190-w (2019).

[46] Perdikaris, P. & Karniadakis, G. E. Model inversion via multi-fidelity Bayesian optimization: a new paradigm for parameter estimation in haemodynamics, and beyond. Journal of the Royal Society Interface 13(118), 20151107 (2016).

[47] Perdikaris, P., Raissi, M., Damianou, A., Lawrence, N. D. & Karniadakis, G. E. Nonlinear information fusion algorithms for robust multi-fidelity modeling. Proceedings of the Royal Society A / Mathematical, Physical and Engineering Sciences 473, 0751 (2017).

[48] Quade, M., Abel, M., Kutz, J. N. & Brunton, S. L. Sparse identification of nonlinear dynamics for rapid model recovery. Chaos 28, 063116 (2018).

[49] Raissi, M., Perdikaris, P., & Karniadakis, G.E. Inferring solutions of differential equations using noisy multi-fidelity data. Journal of Computational Physics 335, 736–746 (2017a).

[50] Raissi, M., Perdikaris, P., & Karniadakis, G.E. Machine learning of linear differential equations using Gaussian processes. Journal of Computational Physics 348, 683-693 (2017b).

[51] Raissi, M., Perdikaris, P., & Karniadakis, G.E. Physics informed deep learning (Part I): Data-driven solutions of nonlinear partial differential equations. ArXiv Prepr ArXiv171110561 (2017c).

[52] Raissi, M., Perdikaris, P., & Karniadakis, G.E.. Physics informed deep learning (Part II): Data-driven discovery of nonlinear partial differential equations. ArXiv Prepr ArXiv171110566 (2017d).

[53] Raissi, M. & Karniadakis, G.E. Hidden physics models: Machine learning of nonlinear partial differential equations. Journal of Computational Physics 357, 125-141 (2018).

[54] Raissi, M., Yazdani, A., & Karniadakis, G. E. Hidden fluid mechanics: A Navier-Stokes informed deep learning framework for assimilating flow visualization data. arXiv preprint arXiv:1808.04327 (2018).

[55] Raissi, M., Perdikaris, P., & Karniadakis, G. E. Physics-informed neural networks: A deep learning framework for solving forward and inverse problems involving nonlinear partial differential equations. Journal of Computational Physics 378, 686–707 (2019).

[56] Rajan, K. & Sreejith, C. Retinal image processing and classification using convolutional neural networks. In: International Conference on ISMAC in Computational Vision and Bio-Engineering, 1271-1280, Springer (2018).

[57] Rhodes, S. J., Knight, G. M., Kirschner, D. E., White, R. G. & Evans, T. G.. Dose finding for new vaccines: The role for immunostimulation/immunodynamic modelling, Journal of Theoretical Biology 465, 51-55 (2019).

[58] Rudy, S. H., Brunton, S. L., Proctor, J. L. & Kutz, J. N. Data-driven discovery of partial differential equations. Science Advances 3(4), e1602614 (2017).

[59] Sahli Costabal, F., Choy, J. S., Sack, K. L., Guccione, J. M., Kassab, G. S. & Kuhl, E. Multiscale characterization of heart failure. Acta Biomaterialia 86, 66-76 (2019).

[60] Sahli Costabal, F., Matsuno, K., Yao, J., Perdikaris, P. & Kuhl, E. Machine learning in drug development: Characterizing the effect of 30 drugs on the QT interval using Gaussian process regression, sensitivity analysis, and uncertainty quantification. Computer Methods in Applied Mechanics and Engineering 348, 313-333 (2019a).

[61] Sahli Costabal, F., Perdikaris, P., Kuhl, E. & Hurtado, D. E. Multi-fidelity classification using Gaussian processes: accelerating the prediction of large-scale computational models. Computer Methods in Applied Mechanics and Engineering 357:112602 (2019b).

[62] Sahli Costabal, F., Seo, K., Ashley, E., & Kuhl, E. Classifying drugs by their arrhythmogenic risk using machine learning. bioRxiv doi: 10.1101/545863 (2019c).

[63] Sanchez-Lengeling, B., & Aspuru-Guzik, A. Inverse molecular design using machine learning: Generative models for matter engineering. Science, 361, 360-365 (2018).

[64] Shaked, I., Oberhardt, M. A., Atias, N., Sharan, R. & Ruppin, E. Metabolic network prediction of drug side effects. Cell Systems 2, 209–213 (2018).

[65] Snowden, T. J., van der Graaf, P. H. & Tindall, M. J. Methods of model reduction for large-scale biological systems: A survey of current methods and trends. Bulletin of Mathematical Biology 79(7), 1449–1486 (2017).

[66] Southern, J., Pitt-Francis, J., Whiteley, J., Stokeley, D., Kobashi, H., Nobes, R., Kadooka, Y. & Gavaghan, D. Multi-scale computational modelling in biology and physiology. Progress in Biophysics and Molecular Biology 96, 60–89 (2008).

[67] Tank, A., Covert, I., Foti, N., Shojaie, A., & Fox, E. Neural Granger causality for nonlinear time series. Retrieved from http://arxiv.org/abs/1802.05842 (2018).

[68] Tartakovsky, A. M., Marrero, C. O., Perdikaris, P., Tartakovsky, G. D., & Barajas-Solano, D. Learning parameters and constitutive relationships with physics informed deep neural networks. Retrieved from http://arxiv.org/abs/1808.03398 (2018).

[69] Tartakovsky, G., Tartakovsky, A. M., & Perdikaris, P. Physics informed deep neural networks for learning parameters with non-Gaussian non-stationary statistics. Retrieved from https://ui.adsabs.harvard.edu/abs/2018AGUFM.H21J1791T (2018).

[70] Teichert, G. & Garikipati, K. Machine learning materials physics: Surrogate optimization and multi-fidelity algorithms predict precipitate morphology in an alternative to phase field dynamics. Computer Methods in Applied Mechanics and Engineering 344, 666-693 (2019).

[71] Teichert, G. H., Natarajan, A. R., Van der Ven, A., & Garikipati, K. Machine learning materials physics: Integrable deep neural networks enable scale bridging by learning free energy functions. Computer Methods in Applied Mechanics and Engineering 353, 201-216 (2019).

[72] Topol, E. J. Deep medicine: how artificial intelligence can make healthcare human again. Hachette Book Group, New York (2019).

[73] Topol, E. J. High-performance medicine: the convergence of human and artificial intelligence. Nature Medicine 25, 44-56 (2019a).

[74] Topol, E. J. Deep learning detects impending organ injury. Nature 572, 36-37 (2019b).

[75] van den Bedem, H. & Fraser, J. Integrative, dynamic structural biology at atomic resolution—It's about time. Nature Methods 12(4), 307-318 (2015).

[76] Vu, M. A. T., Adali, T., Ba, D., Buzsaki, G., Carlson, D., Heller, K., Liston, C., Rudin, C., Sohal, V. S., Widge, A. S., Maybert, H. S., Sapiro, G. & Dzirasa, K. A Shared vision for machine learning in neuroscience. The Journal of Neuroscience: The Official Journal of the Society for Neuroscience 38(7), 1601–1607 (2018).

[77] Walpole, J., Papin, J. A., Peirce, S. M. Multiscale computational models of complex biological systems. Annual Review of Biomedical Engineering 15, 137-154 (2013).

[78] Wang, Z., Huan, X. & Garikipati, K. Variational system identification of the partial differential equations governing the physics of pattern-formation: Inference under varying fidelity and noise. Computer Methods in Applied Mechanics and Engineering (2019).

[79] Weickenmeier, J., Jucker, M., Goriely, A. & Kuhl, E. A physics-based model explains the prion-like features of neurodegeneration in Alzheimer's disease, Parkinson's disease, and amyotrophic lateral sclerosis. Journal of the Mechanics and Physics of Solids 124, 264-281 (2019).

[80] White, R., Peng, G. & Demir, S. Multiscale modeling of biomedical, biological, and behavioral systems. IEEE Eng Med Biol Mag 28, 12-13, (2009).

[81] Xu, M., Papageorgiou, D. P., Abidi, S. Z., Dao, M., Zhao, H., & Karniadakis, G. E. A deep convolutional neural network for classification of red blood cells in sickle cell anemia. PLoS Comp. Bio. 13, e1005746 (2017).

[82] Yang, Z., Yabansu, Y.C., Al-Bahrani, R., Liao, W.K., Choudhary, A.N., Kalidindi, S.R., & Agrawal. Deep learning approaches for mining structure-property linkages in high contrast composites from simulation datasets. Comp Mat Sci 151, 278-287 (2018a).

[83] Yang, L., Zhang, D. & Karniadakis, G.E. Physics-informed generative adversarial networks for stochastic differnetial equations. ArXiv:181102033 [StatML] (2018).

[84] Yang, Y. & Perdikaris, P. Adversarial uncertainty quantification in physics-informed neural networks. Journal of Computational Physics, accepted (2019).

[85] Zhao, L., Li, Z., Caswell, B., Ouyang, J. & Karniadakis, G. E. Active learning of constitutive relation from mesoscopic dynamics for macroscopic modeling of non-Newtonian flows. Journal of Computational Physics 363, 116-127 (2018).